%% file: ms.tex
\documentclass{emulateapj}

\newcommand{\msun} {$M_{\odot}$}
\newcommand{\Te} {$T_{\rm eff}$}
\newcommand{\logg} {$\log g$}

\begin{document}

\title{The Discovery of the Most Metal-Rich White Dwarf: Composition of 
a Tidally Disrupted Extrasolar Dwarf Planet}

\author{P. Dufour\altaffilmark{1}, M. Kilic\altaffilmark{2}, G. Fontaine\altaffilmark{1}, P. Bergeron\altaffilmark{1}, F.-R. Lachapelle\altaffilmark{1}, S. J. Kleinman\altaffilmark{3}, S. K. Leggett\altaffilmark{3}}
\email{dufourpa@astro.umontreal.ca}

\altaffiltext{1}{D\'epartement de Physique, Universit\'e de
  Montr\'eal, Montr\'eal, QC H3C 3J7, Canada}

\altaffiltext{2}{Smithsonian Astrophysical Observatory, 60 Garden St.,
  Cambridge, MA 02138, USA}

\altaffiltext{3}{Gemini Observatory, Northern Operations Center, Hilo,
  HI 96720, USA}

\begin{abstract}

  Cool white dwarf stars are usually found to have an outer atmosphere
  that is practically pure in hydrogen or helium.  However, a small
  fraction have traces of heavy elements that must originate from the
  accretion of extrinsic material, most probably circumstellar
  matter. Upon examining thousands of Sloan Digital Sky Survey
  spectra, we discovered that the helium-atmosphere white dwarf SDSS
  J073842.56+183509.6 shows the most severe metal pollution ever seen
  in the outermost layers of such stars. We present here a
  quantitative analysis of this exciting star by combining high S/N
  follow-up spectroscopic and photometric observations with model
  atmospheres and evolutionary models. We determine the global
  structural properties of our target star, as well as the abundances
  of the most significant pollutants in its atmosphere, i.e., H, O,
  Na, Mg, Si, Ca, and Fe. The relative abundances of these elements
  imply that the source of the accreted material has a composition
  similar to that of Bulk Earth. We also report the signature of a
  circumstellar disk revealed through a large infrared excess in $JHK$
  photometry.  Combined with our inferred estimate of the mass of the
  accreted material, this strongly suggests that we are witnessing the
  remains of a tidally disrupted extrasolar body that was as large as
  Ceres.
 
\end{abstract}


\section{INTRODUCTION}

The vast majority of stars, some 97\% of them, start their lives with
masses less than about 8 \msun. After going through mass loss episodes
in the red giant phases, these stars ultimately run out of
thermonuclear fuel and end up as compact objects known as white
dwarfs. The white dwarf stars are characterized by masses around
0.6 \msun~ and dimensions comparable to that of the Earth, and they
slowly cool off over periods of billions of years by radiating away
the remaining thermal energy left in their core. Most of them are
composed mainly --- more than $99\%$ of their mass --- of carbon and
oxygen, the products of hydrogen and helium nuclear burning. However,
the processes of nuclear fusion and mass loss do not destroy or
evaporate 100$\%$ of the hydrogen and helium that were initially
present in the star at birth. Since the surface gravity of a white
dwarf is extremely high (log $g \sim$ 8), light elements such as
hydrogen and helium float to the surface in enough quantities to form
an optically thick photosphere, while heavier elements sink rapidly
out of sight. It is this efficient gravitational separation mechanism
that is responsible for the high purity of the hydrogen or helium
found in the atmospheres of most white dwarfs.

Traces of heavy elements are sometimes detected spectroscopically in
the atmospheres of cool white dwarfs where radiative levitation or
residual stellar winds no longer operate. Since their gravitational
settling timescales are much shorter than the evolutionary cooling
time, these elements cannot be primordial and must have accreted
relatively recently \citep[metals sink on timescales of at most a few
million years, while white dwarfs cool for several billion years,
][]{paquette}.

Passage through clouds in the interstellar medium (ISM) was, until
recently, the commonly accepted scenario to explain the atmospheric
pollution of these white dwarfs. However, two weaknesses in this
hypothesis have been 1) the amount of hydrogen accreted onto
helium-atmosphere stars is found to be several orders of magnitude
lower than that of metals \citep{wolff,dufour07} despite the fact that
the ISM is mostly constituted of hydrogen, and 2) investigations of
the galactic positions and kinematics of white dwarfs with traces of
metals all failed to show any evidence of an interaction with the ISM
\citep{aannestad,kilic07,farihi10}.

The last decade has seen the rise of a new paradigm as observations at
infrared wavelengths revealed the presence of dusty disks orbiting the
most metal-rich white dwarfs
\citep{becklin,kilic05,kilic06,vonhippel,jura07a,jura07b,farihi09}. It
now appears, in all likelihood, that the heavy elements in most, if
not all, polluted cool white dwarfs are accreted from this orbiting
reservoir whose origin is explained by the tidal disruption of one or
many orbitally perturbed asteroids
\citep{debes,jura03,jura06,jura08}. This alternative model also
naturally explains the lack of correlation with the ISM and the small
amount of hydrogen in the accreting material.

Traces of metals in the atmospheres of cool helium-rich white dwarfs
are usually revealed only by the presence of the broad Ca II H and K
absorption lines in the optical \citep[see][and references
therein]{dufour07}, and very few stars show other heavy elements
\citep [note that observations in the ultraviolet part of the
electromagnetic spectrum, unfortunately available only for a few
objects, see][reveal traces of a few more elements, with the strongest
transitions being due to C, Mg, Si and Fe]{wolff,friedrich,koester00}.

In a few rare cases, however, lines from a handful of elements can be
observed in the optical. GD 362 and GD 40 (see Figure 1) are two of
the most heavily polluted helium-rich white dwarfs currently
known. Both of these stars also possess infrared emitting debris disks
\citep{kilic05,jura07a,jura07b} which are believed to be the remains
of a single tidally disrupted minor body
\citep{jura06,jura08}. Pioneering analyses of these two stars, based
on Keck high resolution observations, have led to the determination of
the relative abundances of 16 and 9 elements respectively
\citep{zuckerman,klein}. The abundance patterns for these two objects
are remarkably consistent with the idea that the source of the
material is asteroids with a composition similar to that of Bulk Earth
\citep{zuckerman,klein}. Studies of heavily polluted white dwarfs can
thus offer a unique opportunity to learn about the chemistry of rocky
extrasolar planet, asteroid belts, and their subsequent dynamical
evolution following the red giant phases. As such, it is of utmost
importance to increase the sample of white dwarfs with a comprehensive
set of measured element abundances.

\begin{figure}[!ht]
\plotone{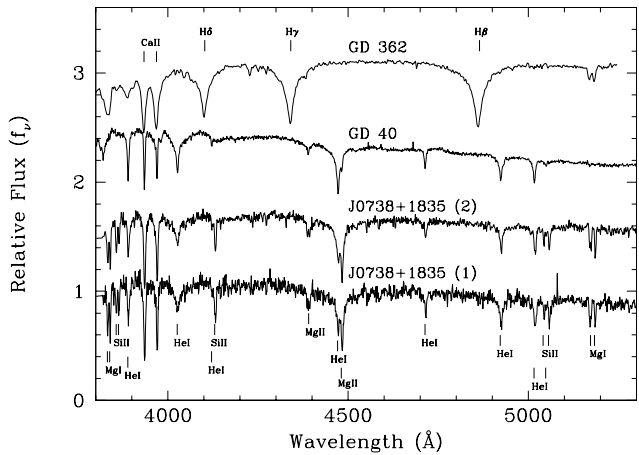}
\caption{Medium resolution spectra of
    J0738+1835, GD 40, and GD 362. The two spectra of J0738+1835 and
    that of GD 40 are from SDSS, while the spectrum of GD 362 is taken
    from Gianninas et al. (2004).}
\end{figure}

In what follows, we present a detailed analysis of SDSS
J073842.56+183509.6 (hereafter J0738+1835), the most metal-rich white
dwarf discovered so far. In \S~\ref{observation}, we describe the
observations. Our detailed analysis follows in \S~\ref{analysis}, and
the results are discussed and summarized in \S~\ref{conclusion}.

\section{TARGET SELECTION AND OBSERVATIONS}\label{observation}

Recently, thanks to the Sloan Digital Sky Survey (SDSS), the number of
spectroscopically identified white dwarfs has increased spectacularly
\citep[more than tenfold, and still counting, Kleinman et al. 2010, in
preparation, see also][]{Eisenstein06}, thus providing a large sample
of potential candidates amenable to detailed chemical analysis. In
order to find such candidates, we selected from the SDSS 7th data
release all spectroscopic objects with $g$ $<$ 19.5 mag that fall
within the color space for white dwarf stars. These spectra were then
visually inspected and classified into the various known white dwarf
spectral types. It is while completing this exercise that the
remarkable spectrum of SDSS J073842.56+183509.6 came to our attention
(see Figure 1). J0738+1835 is a DBZ white dwarf that shows not only
the usual Ca II H and K lines, but also several exceptionally strong
lines of O, Fe, Mg and Si. Remarkably, note that the Mg II
$\lambda$4481 line is deeper than He I $\lambda$4471, a unique feature
among known polluted white dwarfs. The severity of the metallic
pollution of J0738+1835 is particularly striking when its spectrum is
compared to those of GD 362 and GD 40, two objects already known to be
among the most polluted helium-rich white dwarfs currently identified
(see Figure 1). Hence, J0738+1835 is a white dwarf star that offers
great potential for a precise measurement of circumstellar material
composition, and it is only the third helium-rich white dwarf (after
GD 362 and GD 40) for which such a detailed analysis is possible from
optical spectroscopy.

The SDSS photometric magnitudes, in the $ugriz$ system, are $u$ =
17.527 $\pm$ 0.014, $g$ = 17.577 $\pm$ 0.009, $r$ = 17.822 $\pm$
0.009, $i$ = 18.077 $\pm$ 0.010 and $z$ = 18.321 $\pm$ 0.025,
suggesting an effective temperature in the 13,000-15,000 K range,
depending on the amount of reddenning that is assumed. Two spectra of
this object, covering the 3800-9200 \AA~ region at a resolution of
$\sim$ 3\AA~FWHM are available from the SDSS archive. The
SDSS-Modified Julian Date, plate and fiber ID number for these two
spectra are respectively 53431-2054-346 and 54495-2890-354. These two
original SDSS spectra of J0738+1835 presented in Figure 1 are too
noisy for a precise determination of the atmospheric parameters
(effective temperature, surface gravity, and abundances of the various
elements). We thus, as a first step, secured a new medium resolution
high S/N spectrum on UT 2009, November 19, using the 6.5 m MMT
telescope on Mount Hopkins, Arizona, equipped with the Blue Channel
Spectrograph. We used a 1$\arcsec$ slit and the 500 line mm$^{-1}$
grating in first order to obtain spectra with a wavelength coverage of
3500 $-$ 6630 \AA\ and a dispersion of 1.2 \AA~ per pixel. All spectra
were obtained at the parallactic angle. We used He-Ne-Ar comparison
lamp exposures and blue spectrophotometric standards \citep{massey88}
for wavelength and flux calibration, respectively.

In order to verify if an emitting debris disk is present (see below),
infrared observations are needed. We obtained $JHK$ photometry of
J0738+1835 using the Near Infra-Red Imager and Spectrometer (NIRI) on
Gemini-North. These observations were performed on 2010, January 03 as
part of the Director's discretionary time program GN-2009B-DD-8.  We
used a 13 position dither pattern with 21-30 s exposures. We do not
detect any sources above the background within 10" of the target. We
used the Gemini/NIRI package in IRAF to reduce the data and the UKIRT
faint standards \citep{leggett06} to calibrate the photometry. The
derived magnitudes, in the Mauna Kea photometric system, are $J$ =
17.965 $\pm$ 0.031, $H$ = 17.758 $\pm$ 0.033 and $K$ = 17.327 $\pm$
0.034 (or 0.1034 $\pm$ 0.0031 mJy, 0.0791 $\pm$ 0.0026 mJy and 0.0752
$\pm$ 0.0025 mJy respectively).

\section{DETAILED ANALYSIS}\label{analysis}

\subsection{Model Atmospheres and Fitting Technique}

The standard technique to determine the atmospheric parameters for
helium-rich white dwarfs is to perform a $\chi^2$ minimization of the
normalized He I line profiles using a grid of synthetic spectra.
However, in the case of J0738+1835, most of the He I lines are
contaminated by heavy element absorption and finding good
normalization points in the continuum is difficult due to the presence
of numerous metallic lines. Moreover, metallic absorptions have a
significant impact on the thermodynamic structure of the models,
making estimations of the effective temperature and gravity from the
He I lines alone unreliable. For example, in Figure 2, we show the
influence of metals on the temperature and pressure structure of a
\logg = 8, \Te = 14,000 K model atmosphere. This effect is explained
mostly by the flux redistribution due to line blanketing in the UV. We
find that the iron lines provide the dominant contribution in that
regard: for example, the thermodynamic structure of a model calculated
without the iron line opacity is much closer to that of a pure helium
model.

\begin{figure}[!ht]
\plotone{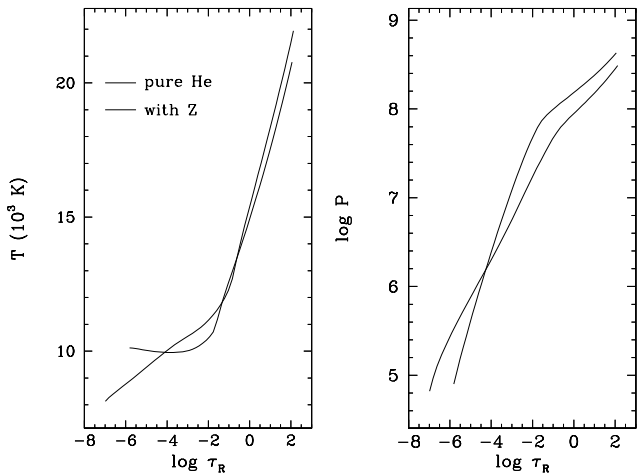}
\caption{Temperature and pressure as a function
    of the Rosseland optical depth for white dwarf model atmosphere
    with \Te = 14,000 K and \logg = 8.0. Thick line is for a pure
    helium atmosphere while the thin line is for a helium-rich
    atmosphere that includes traces of heavy elements (log (Ca/He) =
    -7.0 with the abundances of other elements relative to Ca set as
    described in the text).}
\end{figure}

In this connection, it is interesting to note that our own fit of an
optical SDSS spectrum of GD 40 \citep[another DBZ star analysed in
depth by][] {klein}, using only the He I lines as in the standard
approach and assuming \logg = 8.0 and a pure helium atmosphere, gives
\Te = 15,180 K, very similar to the 15,300 K of \citet{voss07}. In
contrast, a fit with a grid that includes heavy elements yields an
effective temperature up to 1000 K cooler, depending on the exact
amount of metals assumed. This should serve as a warning concerning
the use of pure helium atmospheres for the analysis of such stars.

Given those circumstances, we prefer to fit simultaneously the He I
lines and the numerous iron lines. This is performed by fitting the
whole spectrum with the solid angle, effective temperature, surface
gravity, and the abundance of iron as free parameters. In order to
take into account, to first order, the effect of other heavy elements,
we simply assume that the proportions of all these elements relative
to iron are the same as CI chondrites \citep{lodders}. This is a
reasonable assumption, as demonstrated by the study of GD 362 and GD
40 \citep{zuckerman,klein}, and has the advantage of minimizing the
consequences of unaccounted absorption from heavy elements other than
iron. In other words, it is better to include absorption from
additional metals rather than neglect their contribution completely.
This was chosen over a time-consuming strategy involving navigation in
a N(Z)+2 dimensinoal parameter space (N(Z) elements, effective
temperature, and surface gravity).

We use state-of-the-art model atmospheres based on a code similar to
that described in \citet{dufour07} and \citet{dufourphd}. We use
calcium as a reference element and set the abundances of all elements
with the proportions given by \citep{lodders}. This assumption needs
to be verified a posteriori (see below). Our model grid covers a range
from \Te = 13,000 to 16,000 K in steps of 500 K, \logg = 7.5 to 9.0 in
steps of 0.5 dex, and from log(Ca/He) = $-$5.5 to $-$7.0 in steps of
0.5 dex. Additional models with traces of hydrogen have also been
calculated. All models are calculated with the standard ML2
parametrization of the convective efficiency, but with a mixing length
$\alpha =1.25$ \citep{beauchamp}.

We first start our fitting procedure by assuming the canonical value
\logg = 8.0, but other values of the surface gravity ultimately had
to be explored (see below). We evaluated, as explained above, the
effective temperature and the iron abundance by fitting the optical
spectra. If the lines of a given heavy element are found to be in
disagreement with the observations, we readjust its abundance
accordingly and a new model is recalculated with the new abundances in
a self-consistent way. It is found that small variations of the
abundances of elements other than iron have only a marginal impact on
the thermodynamic structure of the star. Our inferred parameters are
thus not sensitive to our initial assumptions.

However, we noticed that it was not possible to get a satisfactory
simultaneous fit to the Mg I and Mg II lines under the assumption of
log $g$ = 8.0. For example, a good fit to the various Mg II lines
could be achieved only at the expense of a very bad fit to the Mg I
doublet ($\lambda\lambda$ 3832, 3838), the latter being not strong
enough for the assumed atmospheric parameters. This could have been an
indication that the effective temperature was overestimated, but the
much lower value needed to reconcile the Mg I and Mg II line strengths
was clearly incompatible with the He I and Fe I/Fe II line
profiles. Our first idea was that the thermodynamic structure of our
model was wrong due to one of the assumptions about the composition,
but models calculated by including only the observed elements were
essentially identical to the ones with all the elements included. We
also explored the parameter space to see if another combination of
abundances and effective temperature could solve that problem, but no
solutions were found.

Another way to favor the formation of Mg I relative to Mg II is to
increase the atmospheric pressure. This can be achieved if the surface
gravity is allowed to increase in our fitting procedure. In a manner
similar to that described above, we find that the Mg I and Mg II lines
can be fitted simultaneously (as well as the He I, Fe I, and Fe II
lines) if we increase the surface gravity to \logg = 8.5. Slightly
larger/lower values of \logg~ (compensated with an accordingly
larger/lower effective temperature) can also accommodate
simultaneously the Mg I and Mg II lines, but then the He I lines fit
is not as good. Our derived mass determination thus relies heavily on
the magnesium lines and the somewhat uncertain He I line profiles
(usually attributed to the treatment of van der Waals broadening).

Trends of higher mass at low effective temperature for DB stars has
raised concerns in the literature about the validity of the derived
masses from He I lines. However, as clearly shown in \citet[][see
their Figure 7]{limoges}, it seems that the spread in mass is probably
real after all. Since a higher than average surface gravity best fits
both the He I lines and the Mg lines, we have a high confidence that
J0738+1835 is indeed relatively massive. The debate will be settled
when a trigonometric parallax measurement and better broadening
theories become available, but until that happens, we consider our
solution as the best compromise.

There appears to be large discrepancies in the predicted line strength
of a few iron lines, the most noticeable being that of Fe II
$\lambda$4173.46 and near 5200 \AA (see Figure 3 and 4). These
discrepancies cannot be eliminated even by varying significantly from
our optimal solution the surface gravity and the effective temperature
of our model. It is most probable that this is a manifestation of the
uncertainties in the atomic data, $\log$ gf and Stark broadening
parameters in particular, used in our calculation (we used the well
known Kurucz linelists). We calculated a posteriori a synthetic
spectrum using the Vienna Atomic Line Database (VALD) lists. Using
these data helps to partly reduce the discrepancies for a few iron
lines (for example, the depth of the Fe II $\lambda$4173.46 is reduced
by half) but does not alter the abundance determinations for any
elements on average. Since our fit is based on numerous iron lines,
these few discrepant lines have a relatively weak weight on the iron
abundance determination and we present our final solution using the
Kurucz linelist for consistency with the rest of our analysis.

\begin{figure}[!ht]
\plotone{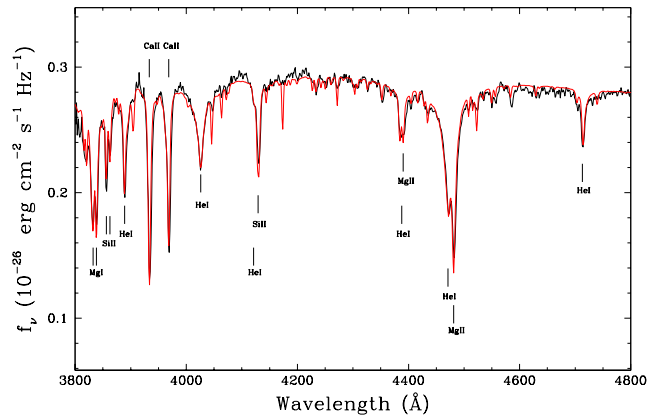}
\caption{MMT spectrum (in black) and our best
    fit model (red line). The strongest lines are indicated by tick
    marks. All the other numerous non-identified lines are from
    iron. They have not been marked for clarity. In order to take into
    account uncertainties on the slope of the continuum due to
    extinction and flux calibration, a small linear and quadratic term
    were also included in the fitting procedure.}
\end{figure}

\begin{figure}[!ht]
\plotone{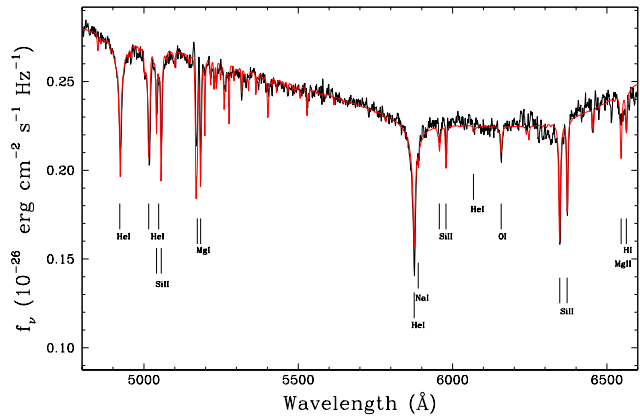}
\caption{Same as Figure 3, but for the red
    part of the spectrum. Note that due to second order contamination
    longward of 5900 \AA, a second-order polynomial fit has been
    applied to the model spectrum continuum.}
\end{figure}

We note that the assumption on the abundance of heavy elements
described above also predicts weak lines of sodium (the Na D lines)
that are blended with the He I $\lambda$5876 line. There appear to be a
feature in the red wing of the He I line that is consistent with the
expected lines strength but given the signal-to-noise ratio of our
observations and the weakness of the presumable lines, we shy away
from claiming a detection. Two weak lines of Cr and Ti are also barely
predicted at a level comparable to the signal-to-noise ratio of our
observations. Future observations with higher signal-to-noise and
higher resolution (as well as in other part of the electromagnetic
spectrum) should eventually confirm their presence.

Our final solution is presented in in Table~1 while our best fit model
spectra are shown in Figures 3 and 4. Uncertainties obtained from the
covariance matrix are unrealistically small. These represent only
internal errors and thus underestimate the real uncertainties. We thus
evaluate the uncertainties on the model parameters by calculating a
series of models with various parameters near our optimal
solution. The constraint from fitting simultaneously the Mg I, Mg II
and He I lines dominates the errors on the effective temperature and
surface gravity, which we evaluate to be about 300 K and 0.2 dex
respectively. The uncertainties on the element abundances must also
take into account these uncertainties. In a similar manner to that
described in \citet{klein}, we varied the model \Te~ and \logg~ by
their uncertainties, one at a time, to evaluate the corresponding
errors on the various element abundances. The final uncertainties on
the abundances presented in Table~1 are the sum of the errors added in
quadrature. Those uncertainties are then properly propagated for the
determination of the errors on the cooling age, luminosity, mass,
radius and distance of the star (obtained by determining the solid
angle from the $ugriz$ photometry). The evolutionary models used are
similar to those described in \citep{FBB01} but with C/O cores, and
thickness of the helium and hydrogen layers of respectively q(He) =
10$^{-2}$ and q(H) = 10$^{-10}$, which are representative of
helium-rich atmosphere white dwarfs.

\input{tab1}

Finally, we calculated a posteriori a model with the final abundances
of only the observed elements. We find that the resulting synthetic
spectrum is practically identical to the one with all elements
included. This is not surprising given that Fe, Mg, Si, O, and Ca account
for more than $95\%$ of the contaminants by mass for Bulk Earth
\citep[a similar proportion was also found for GD 40,][]{klein}. A
similar proportion will probably be retrieved for J0738+1835 when high
resolution observations become available (ultraviolet observations
should also be helpful).

\subsection{Convection Zone Models}

In order to evaluate the total amount of heavy material mixed in the
outer layers of J0738+1835, we need to know the mass of the helium
convection zone into which the metals are diluted. Estimates of this
mass can readily be obtained using the Montr\'eal white dwarf building
codes that were developed over the years \citep[see,
e.g.,][]{BF94,FBB01}, and which have been maintained at the
state-of-the-art level. One possible option offered by these codes is
to compute complete, but static, stellar structures with a luminosity
profile closely following the mass profile, as appropriate for cool
white dwarfs shining through the loss of thermal energy. This is what
we adopted here.

To ease the comparison with the spectroscopic observations, we further
selected the option of fixing the surface gravity and the effective
temperature (as well as the compositional stratification and the core
composition) in these stellar model calculations. We also assumed
relatively thick helium envelopes in these models, log $(1 - M_{\rm
  He}/M_*) = -3.0$, much thicker than the superficial outer convection
zone due to helium partial ionization. Since the presence of heavy
elements does affect somewhat the extent of the convection zone,
models with an envelope composition consisting of helium and
homogeneous traces of metals were considered along with more standard
models having pure helium envelopes. We retained the results for a
metallicity of Z = 0.001 in what follows.

We investigated also how $M_{\rm He}/M_*$ depends on the assumed
convective efficiency. This is because, as was first shown by
\citet{ber95}, the calibration of the mixing-length derived in the
atmospheric layers (by comparing optical and UV observations) does not
apply at the base of the convection zone, where the flux of settling
heavy elements determines the observable abundances. In that case, the
calibration is best obtained by comparing the effective temperatures
at the blue edge of an instability strip as inferred from nonadiabatic
pulsation calculations with the values derived from spectroscopy. In
the case of pulsating H-rich (DA) white dwarfs, \citet{ber95} found
that the convective efficiency increases with depth, from the
atmospheric layers to the bottom of the convection zone, so that a
variable mixing length is required to account for both the
spectroscopic and pulsational measurements. Unfortunately, the
calibration based on pulsational properties is still uncertain,
particularly for the He-atmosphere white dwarfs \citep [see,
e.g.,][]{FB08}, so we decided to consider three different plausible
versions of the mixing-length theory for constraining the mass of the
outer convection zone in J0738+1835.

The results of our calculations are presented in Table~2 for values of
log $g$ and $T_{\rm eff}$ in the vicinity of those found
spectroscopically for J0738+1835. Our calculations indicate that the
mass of the helium convection zone in that regime is, fortunately,
only weakly dependent on the effective temperature and on the
convective theory flavor used in the stellar models. However, a strong
dependency on the surface gravity is found, $M_{\rm He}/M_*$
decreasing significantly as the surface gravity increases. Using our
atmospheric parameters, we estimate that the mass of the helium
convection zone lies somewhere between $10^{-5.70}M_*$ and
$10^{-6.75}M_*$. Correspondingly, the lower limits on the mass of the
polluting body (i.e. the sum of the masses of the detected elements in
Table~1) are between $2.7 \times 10^{23}$ g and $2.4 \times 10^{24}$
g, with a value of $ 4.3\times 10^{23}$ g corresponding to \logg =
8.5.

\input{tab2}

This is close to the mass of Ceres ($9.4 \times 10^{23}$ g) and is
about an order of magnitude more than the amount of metals found in
the convection zones of GD 40 \citep{klein} and GD 362
\citep{koester09} ($3.6 \times 10^{22}$ g and $1.8 \times 10^{22}$ g,
respectively), and comparable to the amount found in HS 2253+8023
\citep{jura06} which is, however, based solely on the derived Fe
abundance from an $IUE$ spectrum \citep{friedrich}.  Because an
unknown period of time has elapsed since the accretion process has
started, we do not know what fraction of the material has already sunk
out of sight below the base of the convection zone and, therefore,
this estimate for the total mass of heavy elements represents a lower
limit on the mass of the body responsible for this large
pollution. Given this, and the fact that another yet unknown quantity
of material is still orbiting the star in the form of a debris disk
(see below), it is safe to presume that the object that was tidally
destroyed by J0738+1835 was at least as large as the dwarf planet
Ceres.

This lower limit is sensitive to the determined value of the surface
gravity.  It will thus be of utmost importance to determine the
surface gravity with greater accuracy than is currently possible in
order to precisely constrain the mass of the asteroid/dwarf planet
that causes the pollution in J0738+1835. A good trigonometric parallax
measurement should help to reduce this uncertainty significantly.

\subsection{Infrared Photometry and Disk Model}

The next step was to determine if an accretion disk is still present
around J0738+1835, which would be revealed by an infrared excess. We
thus obtained infrared $JHK$ photometry with the 8 m Gemini North
telescope in Hawaii. Our observations, presented in Figure 5, clearly
reveal a near-IR excess. 

\begin{figure}[!ht]
\plotone{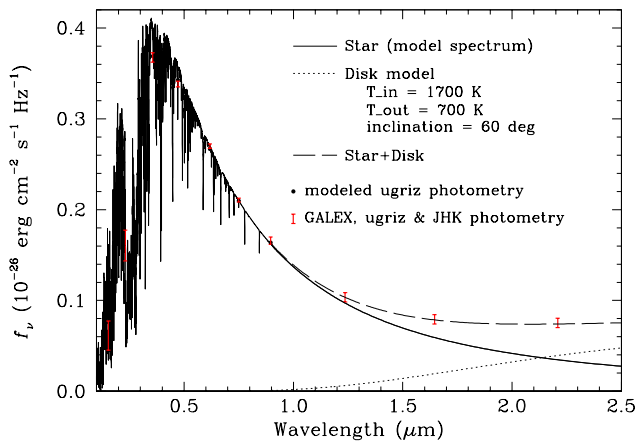}
\caption{Photometric measurements in the GALEX, 
    $ugriz$ and $JHK$ bands compared to models of the star and the
    debris disk.}
\end{figure}

Given the fact that most heavily polluted white dwarfs have debris
disks, our target is likely to have a disk as well. If the
near-infrared excess is due to a companion star, this star would have
$M_{\rm K}$ = 12.35 mag; an L6 dwarf \citep{leggett02}. Such a
companion would have $H-K$ = 0.6-0.9 mag. The near-infrared excess
around our target (after subtracting the contribution from the stellar
photosphere) has a $H-K$ color of 1.33 mag, incompatible with an L6
dwarf or any other M, L, and T dwarfs studied by
\citet{leggett02}. Therefore, the infrared excess around our target is
best explained by the presence of a dusty debris disk rather than a
brown dwarf companion.

We fit the infrared excess around J0738+1835 using the optically thick
flat-disk models of \citet{jura03}. Using the effective temperature,
radius, and distance estimates for the white dwarf (see Table~1), we
created a grid of disk models with inner temperatures 1000-1800 K,
outer temperatures 300-900 K (in steps of 100 K), and inclination
angles 15$^\circ$-75$^\circ$ (in steps of 15$^\circ$). The model that
best fits the infrared photometry has $T_{\rm in}=$ 1700 K, $T_{\rm
  out}=$ 700 K, and an inclination angle of 60$^\circ$. However, since
we only have near-IR data, the outer temperature is not well
constrained, and these parameters change depending on the assumed
inclination. For example, if the inclination angle is assumed to be
75$^\circ$, then the best fit model has $T_{\rm in}$=1600 K and
$T_{\rm out}$=300 K. Nevertheless, the near-IR data can be explained
with a reasonable set of parameters for a dusty disk, indicating that
the accretion process is most probably occurring at present. An inner
temperature of 1700 K is somewhat unusual for disks around white
dwarfs. However, the near-infrared flux excess around J0738+1835 is
similar to the excess seen around the white dwarf GD 56
\citep{kilic06}. A flat-disk model with an inner temperature of 1700 K
is required to explain the observed excess around GD 56
\citep{jura07a}. A better fit to the near-infrared data for
GD 56 can be obtained if the disk is substantially warped or puffed up
by the gravitational field of a planet \citep{jura09}. Similar
processes may be responsible for the observed near-infrared flux
distribution of the disk around J0738+1835. Mid-infrared observations
will be useful to check the flat-disk models for this target.

Unfortunately, the total mass of the disk cannot be determined for an
opaque ring. Masses for the dust disks around G29-38 and GD 362 have
been estimated from models of the mid-infrared emission
\citep{reach,becklin}. Based on the choice of optically thin or thick
models, these disks hold $\sim 10^{19}-10^{24}$ g of material. The
disk around J0738+1835 is likely to hold a similar amount of material
as well. Hence, the combined mass of the metals in the surface
convection zone and the dust disk is on the order of 4-14 $\times
10^{23}$ g, equivalent to the mass of a dwarf planet.

\section{DISCUSSION AND CONCLUSION}\label{conclusion}

According to the accretion/diffusion scenario
\citep{dupuisa,dupuisb,dupuisc}, a steady state abundance for a given
heavy element is rapidly reached after the accretion process
starts. Under those circumstances, the ratio of abundances measured
spectroscopically in the photosphere of J0738+1835 can be related,
unlike the case of HS 2253+802 which shows no disk \citep{farihi09},
to that of the incoming disk material.

It is interesting to note that the mass ratios of the most abundant
heavy elements found in J0738+1835, namely O, Mg, Si and Fe, are very
similar to that found in GD 40, which has an atmosphere polluted with
debris coming from a body with an Earth-like composition
\citep{klein}. However, calcium is depleted in J0738+1835 by a factor
of 4, perhaps indicating that mantle/crust differentiation has
occurred before the crust was lost. Moreover, if we assume that the
total amount of hydrogen present in the atmosphere of J0738+1835
originates from the last polluting event only, we can place an upper
limit of $\sim$1$\%$ by mass on the water or ice content of the
polluting body. This is a surprisingly low amount considering that
most asteroids in our solar system contain large quantities of ice on
their surface. This might indicate that the parent body was
sufficiently close to the progenitor red giant star for thermal
processes to have evaporated most of the ice or water initially
present \citep{jura10}. Thus it is possible that the surface of the
orbiting body was significantly altered in the late stages of stellar
evolution, but at this point it is premature to quantify this
effect. Given the uncertainties on the atmospheric parameters, we
shall refrain from speculating further on the true fate of the body
that polluted the atmosphere of J0738+1835. However, when parallax and
higher resolution spectroscopy observations become available,
J0738+1835 will become a perfect testbed for a precise study of not
only the abundances of the polluting extrasolar body, but also of the
effects of processes such as crust differentiation and thermal
heating.

It is worth noting that medium resolution observations of GD 362 and
GD 40 revealed only relatively weak absorptions of Ca, Mg, and Fe in
the optical (Figure 1). It was only when Keck high
resolution spectra became available that a myriad of elements were
uncovered, providing extremely valuable information about the accreted
material.  Given that J0738+1835's medium resolution spectrum is
showing even stronger metallic absorption lines than in the cases of
GD 362 and GD 40, further high resolution observations in the optical
and ultraviolet to study elemental abundances as well as at longer
wavelengths to characterize the debris disk are highly
desirable. These observations will undoubtedly offer an opportunity to
study the composition of an extrasolar dwarf planet with unprecedented
accuracy.
 
\acknowledgements This work was supported in part by NSERC Canada and
FQRNT Qu\'ebec. P.D is a CRAQ postdoctoral fellow, and P.B is a
Cottrell Scholar of Research for Science
Advancement. G.F. acknowledges the contribution of the Canada Research
Chair Program. MK acknowledges support from NASA through the Spitzer
Space Telescope Fellowship Program, under an award from Caltech. Based
on observations obtained at the MMT and Gemini Observatory. The MMT is
a joint facility of the Smithsonian Institution and the University of
Arizona. The Gemini observatory is operated by the Association of
Universities for Research in Astronomy, Inc., under a cooperative
agreement with the NSF on behalf of the Gemini partnership: the
National Science Foundation (United States), the Science and
Technology Facilities Council (United Kingdom), the National Research
Council (Canada), CONICYT (Chile), the Australian Research Council
(Australia), Ministerio da Ciencia, Tecnologia (Brazil) and Ministerio
de Ciencia, Tecnologia e Innovacion Productiva (Argentina).

\end{document}

%% file: tab1.tex
\begin{deluxetable}{lccccc}
\tabletypesize{\scriptsize}
\tablecolumns{11}
\tablewidth{0pt}
\tablecaption{Stellar parameters for SDSS J0738+1835}
\tablehead{Parameter & Value}
\startdata
$T_{\rm eff}$(K)  & 13600 $\pm$ 300    \\
$\log g$         & 8.5 $\pm$ 0.2    \\
$M_{\rm WD}/M_{\odot}$     & 0.907  $\pm$ 0.128  \\
$M_{\rm init}/M_{\odot}$     & 4.4  $\pm$ 1.0$^a$  \\
$R/R_{\odot}$     & 0.00886 $\pm$ 0.0015  \\
$\log L/L_{\odot}$ & $-$2.62 $\pm$ 0.14\\
$D$              & 136 pc $\pm$ 22\\
Cooling Age      & 595 Myr $\pm$ 219\\
$\log$ H/He      &  $-$5.7 $\pm$ 0.3\\
$\log$ O/He      &  $-$4.0 $\pm$ 0.2\\
$\log$ Mg/He     &  $-$4.7 $\pm$ 0.2\\
$\log$ Si/He     &  $-$4.9 $\pm$ 0.2\\
$\log$ Ca/He     &  $-$6.8 $\pm$ 0.3\\
$\log$ Fe/He     &  $-$5.1 $\pm$ 0.3\\
$\log (M_{\rm He}/M_{\star})$ &  $-$6.5 $+$0.8/$-$0.25 \\
\enddata
\tablenotetext{a}{Initial mass of the main sequence progenitor
  calculated using the Initial-Final Mass Relation of
  \citet{williams09}}
\end{deluxetable}

%% file: tab2.tex
 \begin{deluxetable}{llcccc}
 \tabletypesize{\footnotesize}
 \tablecolumns{5}
 \tablewidth{0pt}
 \tablecaption{ Fractional mass of the convection zone 
 (log $M_{\rm He}/M_*$), for various surface gravities,
   effective temperatures, and convective efficiencies}
 \tablehead{
 \colhead{log $g$} &
 \colhead{$T_{\rm eff}$(K)} &
 \colhead{ML2$/{\alpha=0.6}$} &
 \colhead{ML2} &
 \colhead{ML3}}
 \startdata
7.5 & 13000 &$-$4.344 & $-$4.301 & $-$4.259 &\\
    & 14000 &$-$4.604 & $-$4.518 & $-$4.454 &\\
8.0 & 13000 &$-$5.367 & $-$5.324 & $-$5.324 &\\
    & 14000 &$-$5.543 & $-$5.500 & $-$5.457 &\\
8.5 & 13000 &$-$6.447 & $-$6.447 & $-$6.425 &\\
    & 14000 &$-$6.580 & $-$6.558 & $-$6.537 &\\
9.0 & 13000 &$-$7.661 & $-$7.661 & $-$7.661 &\\
    & 14000 &$-$7.729 & $-$7.707 & $-$7.707 &\\

 \enddata
 \end{deluxetable}

%% file: ms.bbl
\begin{thebibliography}{}
\bibitem[Aannestad et al.(1993)]{aannestad} Aannestad, P.~A., Kenyon,
  S.~J., Hammond, G.~L., \& Sion, E.~M.\ 1993, \aj, 105, 1033

\bibitem[Beauchamp et al.(1999)]{beauchamp} Beauchamp, A., Wesemael,
  F., Bergeron, P., Fontaine, G., Saffer, R.~A., Liebert, J., \&
  Brassard, P.\ 1999, \apj, 516, 887

\bibitem[Becklin et al.(2005)]{becklin} Becklin, E.~E., Farihi, J.,
  Jura, M., Song, I., Weinberger, A.~J., \& Zuckerman, B.\ 2005,
  \apjl, 632, L119
 
\bibitem[Bergeron et al.(1995)]{ber95} Bergeron, P., Wesemael, F.,
  Lamontagne, R., Fontaine, G., Saffer, R.~A., \& Allard, N.~F.\ 1995,
  \apj, 449, 258

\bibitem[Brassard \& Fontaine(1994)]{BF94} Brassard, P., \& Fontaine,
  G.\ 1994, IAU Colloq.~147: The Equation of State in Astrophysics,
  560

\bibitem[Debes \& Sigurdsson(2002)]{debes} Debes, J.~H., \&
  Sigurdsson, S.\ 2002, \apj, 572, 556

\bibitem[Dufour et al.(2007)]{dufour07} Dufour, P., et al.\ 2007,
  \apj, 663, 1291

\bibitem[Dufour(2007)]{dufourphd} Dufour, P.\ 2007, Ph.D.~Thesis, Universit\'e de Montr\'eal

\bibitem[Dupuis et al.(1992)]{dupuisa} Dupuis, J., Fontaine, G.,
  Pelletier, C., \& Wesemael, F.\ 1992, \apjs, 82, 505

\bibitem[Dupuis et al.(1993a)]{dupuisb} Dupuis, J., Fontaine, G.,
  Pelletier, C., \& Wesemael, F.\ 1993a, \apjs, 84, 73

\bibitem[Dupuis et al.(1993b)]{dupuisc} Dupuis, J., Fontaine, G., \&
  Wesemael, F.\ 1993b, \apjs, 87, 345

\bibitem[Eisenstein et al.(2006)]{Eisenstein06} Eisenstein,
  D.~J., et al.\ 2006, \apjs, 167, 40

\bibitem[Farihi et al.(2010)]{farihi10} Farihi, J., Barstow, M.~A.,
  Redfield, S., Dufour, P., \& Hambly, N.~C.\ 2010, arXiv:1001.5025

\bibitem[Farihi et al.(2009)]{farihi09} Farihi, J., Jura, M., \&
  Zuckerman, B.\ 2009, \apj, 694, 805

\bibitem[Fontaine \& Brassard(2008)]{FB08} Fontaine, G., \& Brassard,
  P.\ 2008, \pasp, 120, 1043

\bibitem[Fontaine et al.(2001)]{FBB01} Fontaine, G., Brassard, P., \&
  Bergeron, P.\ 2001, \pasp, 113, 409

\bibitem[Friedrich et al.(1999)]{friedrich} Friedrich, S., Koester,
  D., Heber, U., Jeffery, C.~S., \& Reimers, D.\ 1999, \aap, 350, 865

\bibitem[Gianninas et al.(2004)]{gianninas} Gianninas, A., Dufour, P.,
  \& Bergeron, P.\ 2004, \apjl, 617, L57

\bibitem[Jura(2003)]{jura03} Jura, M.\ 2003, \apjl, 584, L91

\bibitem[Jura(2006)]{jura06} Jura, M.\ 2006, \apj, 653, 613

\bibitem[Jura et al.(2007a)]{jura07a} Jura, M., Farihi, J., \&
  Zuckerman, B.\ 2007, \apj, 663, 1285

\bibitem[Jura et al.(2007b)]{jura07b} Jura, M., Farihi, J., Zuckerman,
  B., \& Becklin, E.~E.\ 2007, \aj, 133, 1927

\bibitem[Jura(2008)]{jura08} Jura, M.\ 2008, \aj, 135, 1785

\bibitem[Jura et al.(2009)]{jura09} Jura, M., Farihi, J., 
\& Zuckerman, B.\ 2009, \aj, 137, 3191 

\bibitem[Jura \& Xu(2010)]{jura10} Jura, M., \& Xu, S.\ 2010,
  arXiv:1001.2595

\bibitem[Kilic \& Redfield(2007)]{kilic07} Kilic, M., \& Redfield, S.\
  2007, \apj, 660, 641

\bibitem[Kilic et al.(2005)]{kilic05} Kilic, M., von Hippel, T.,
  Leggett, S.~K., \& Winget, D.~E.\ 2005, \apjl, 632, L115

\bibitem[Kilic et al.(2006)]{kilic06} Kilic, M., von Hippel, T.,
  Leggett, S.~K., \& Winget, D.~E.\ 2006, \apj, 646, 474

\bibitem[Klein et al.(2010)]{klein} Klein, B., Jura, M., Koester, D.,
  Zuckerman, B., \& Melis, C.\ 2010, \apj, 709, 950


\bibitem[Koester \& Wilken(2006)]{koester06} Koester, D., \& Wilken,
  D.\ 2006, \aap, 453, 1051

\bibitem[Koester \& Wolff(2000)]{koester00} Koester, D., \& Wolff, B.\
  2000, \aap, 357, 587

\bibitem[Koester(2009)]{koester09} Koester, D.\ 2009, \aap, 498, 517

\bibitem[Leggett et al.(2002)]{leggett02} Leggett, S.~K., et al.\ 
2002, \apj, 564, 452 

\bibitem[Leggett et al.(2006)]{leggett06} Leggett, S.~K., et al.\
  2006, \mnras, 373, 781

\bibitem[Limoges \& Bergeron(2010)]{limoges} Limoges, M.-M., \&
  Bergeron, P.\ 2010, \apj, 714, 1037

\bibitem[Lodders(2003)]{lodders} Lodders, K.\ 2003, \apj, 591, 1220

\bibitem[Massey et al.(1988)]{massey88} Massey, P., Strobel, K.,
  Barnes, J.~V., \& Anderson, E.\ 1988, \apj, 328, 315

\bibitem[Paquette et al.(1986)]{paquette} Paquette, C., Pelletier, C.,
  Fontaine, G., \& Michaud, G.\ 1986, \apjs, 61, 197

\bibitem[Reach et al.(2005)]{reach} Reach, W.~T., Kuchner, M.~J., von
  Hippel, T., Burrows, A., Mullally, F., Kilic, M., \& Winget, D.~E.\
  2005, \apjl, 635, L161

\bibitem[von Hippel et al.(2007)]{vonhippel} von Hippel, T.,
Kuchner, M.~J., Kilic, M., Mullally, F., \& Reach, W.~T.\ 2007, \apj, 662, 544

\bibitem[Voss et al.(2007)]{voss07} Voss, B., Koester, D., Napiwotzki,
  R., Christlieb, N., \& Reimers, D.\ 2007, \aap, 470, 1079

\bibitem[Williams et al.(2009)]{williams09} Williams, K.~A.,
  Bolte, M., \& Koester, D.\ 2009, \apj, 693, 355

\bibitem[Wolff et al.(2002)]{wolff} Wolff, B., Koester, D., \&
  Liebert, J.\ 2002, \aap, 385, 995

\bibitem[Zuckerman et al.(2007)]{zuckerman} Zuckerman, B., Koester,
  D., Melis, C., Hansen, B.~M., \& Jura, M.\ 2007, \apj, 671, 872


\end{thebibliography}
